\documentstyle[epsfig,aps,prl,epsf,multicol]{revtex}

\begin{document}

%\twocolumn

\hsize\textwidth\columnwidth\hsize
\csname@twocolumnfalse\endcsname

\title{Dynamical fluctuations in mode locking experiments on vortices \\
moving through mesoscopic channels}

\author{N.~Kokubo, R.~Besseling, and P.H.~Kes}

\address{Kamerlingh Onnes Laboratorium, Leiden
University, P.O. Box 9504, 2300 RA Leiden, the Netherlands.}
\date{\today}
\maketitle

\begin{abstract}
We have studied the flow properties of vortices driven through
easy flow mesoscopic channels by means of the mode locking (ML)
technique. We observe a ML jump with large voltage broadening in
the real part of the rf-impedance. Upon approaching the pure dc
flow by reducing the rf amplitude, the ML jump is smeared out via
a divergence of the voltage width. This indicates a large spread
in internal frequencies and lack of temporal coherence in the
dc-driven state.
\end{abstract}
%\twocolumn
\begin{multicols}{2}
%\raggedcolumns
\narrowtext \noindent

\section{Introduction}

One of the powerful techniques for probing the dynamic state of
driven vortex matter in type II superconductors is the mode
locking (ML) experiment. ML is a dynamic resonance between a
superimposed rf drive and the dynamic lattice mode (as expected to
appear when vortices flow coherently in a (random) pinning
environment) excited collectively at the internal (washboard)
frequency $f_{int}=qv/a$ with $q$ integer, $v$ the average dc
velocity and $a$ the lattice spacing of the array along the flow
direction. When $f_{int}$ and the rf drive frequency $f$ are
harmonically related, i.e., $f_{int}=pf$ with $p$ integer, the
resonance appears as steps in the dc current-voltage ($I-V$)
characteristics and also emerges more sensitively as jumps (dips)
in a real (imaginary) part of rf impedance at low rf drives
$I_{rf}<I_{dc}$ \cite{Fiory}\cite{SH}\cite{Kokubo}.

So far, some features of the ML phenomenon have been well studied
in sliding charge density waves in quasi one dimensional
conductors \cite{Gruner}. It turns out that complete ML features
characterized by "perfect" steps in dc $V-I$ curve appear when the
flow is coherent enough, whereas incomplete features like rounded
and broadened steps in the $V-I$ curve appear if the flow becomes
more incoherent. Thus, the ML feature provides information on the
amount of coherency in the flow. In this study, we have employed
the ML technique to study flow of vortices driven through
disordered mesoscopic channels \cite{Kokubo}.

\section{Experimental}

The channel device consists of a strong pinning NbN film
(thickness $d$=50nm) on top of a weakly pinning amorphous
Nb$_{1-x}$Ge$_x$ film ($d$=550nm and $x\approx$0.3). By reactive
ion etching and proper masking, straight channels were etched
leaving a 300nm thick NbGe layer. The width and length of each
channel is 230nm and 300$\mu$m, respectively.  The pinning
potential for vortices inside the channel is dominated by the
interaction with strongly pinned edge vortices which form a
disordered configuration (see the inset to Fig.1). The transport
current is applied perpendicular to the channel and drives the
vortices along the channel. For ML experiments, we sweep the dc
current with a constant superimposed rf current and record both
the in- and out-of-phase part of the rf-voltage as well as the dc
voltage. The sample was immersed in superfluid $^4$He at a
temperature of 1.9K.

\section{Results and discussion}

Because of the confinement of the vortex array in the channels, an
integer number $n$ of rows is expected to form inside each
channel. In a previous study, we demonstrated that the flow
configuration can be related to the voltage $V_{1,1}$ at which the
fundamental ML ($p=q=1$) occurs \cite{Kokubo}. Namely,
\begin{equation}
  V_{1,1}=f\phi_0nN_{ch}
\end{equation}
with $\phi_0$ the flux quantum and $N_{ch}(\approx 200$) the
number of channels.

\begin{figure}
\begin{center}
\epsfig{file=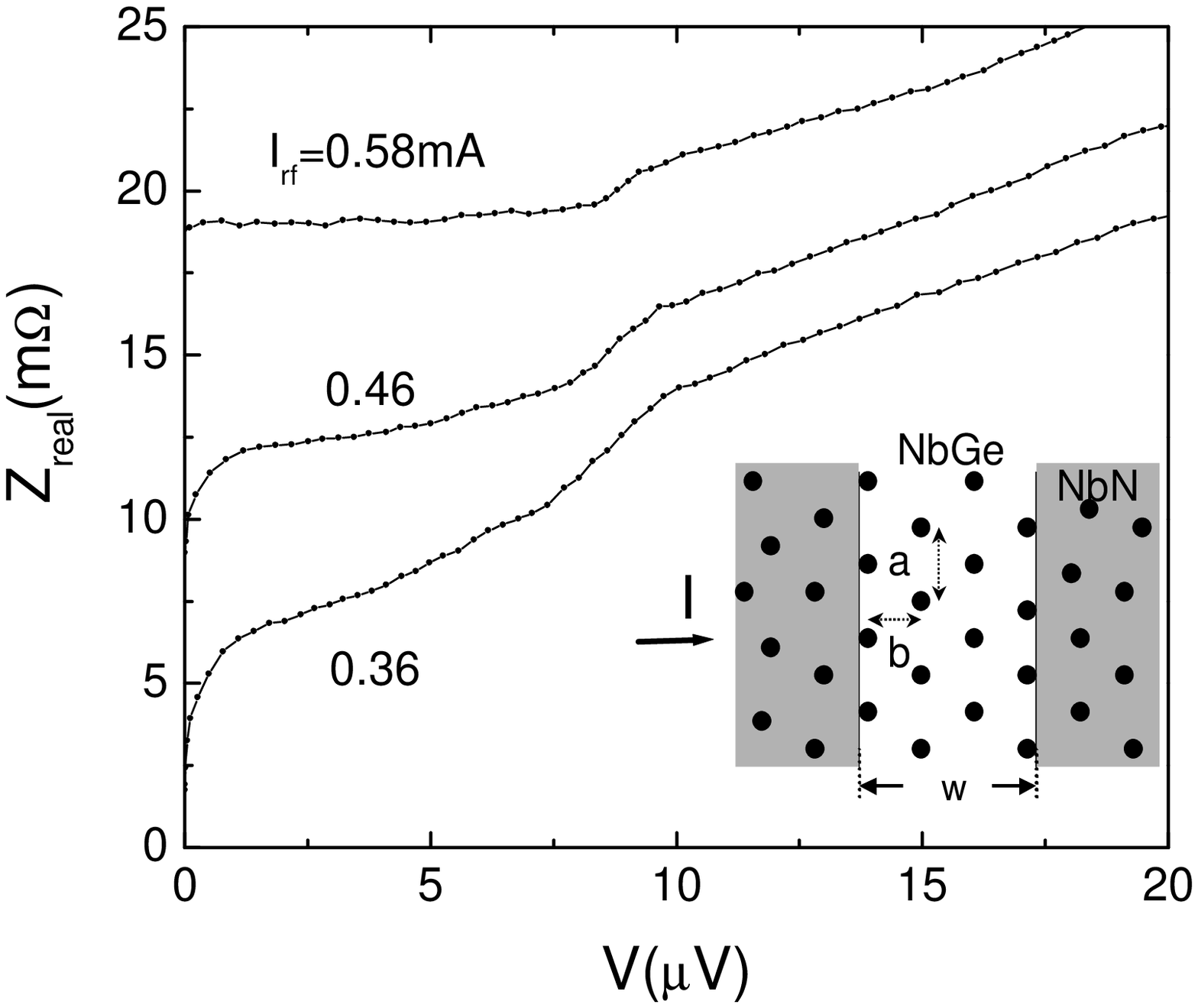, width=5.5cm} \vspace{0.1cm}
\label{Fig.1} \caption{The dc voltage dependence of the real part
of the rf impedance $Z_{real}$ measured by applying 6MHz rf
currents of various amplitudes at 1.9K and 170mT.  The upper two
curves are shifted upwards by 3 and 6 m$\Omega$. The inset shows a
schematic configuration of vortex array around a channel.}
\end{center}
\end{figure}

In Fig.1 we plot the real part of the rf-impedance $Z_{real}$
versus the dc voltage $V$ taken at several magnitudes of rf
currents $I_{rf}$ of 6MHz. We choose a field of 170mT where moving
configurations of $n$=4 rows form in each channel. Around
$V_{1,1}\approx9\mu$V we observe rounded jumps, while below and
above $V_{1,1}$ $Z_{real}$ increases monotonically with $V$. It is
clear that the height and the voltage broadening of the rounded
jump are enhanced when $I_{rf}$ is reduced. For definition of the
jump height $\Delta Z$, we take the derivative of $Z_{real}$-$V$
curve and integrate over the corresponding peak in a differential
plot of $dZ_{real}/dV$ versus $V$. The inset to Fig.2 shows a plot
of $\Delta Z$ as a function of $I_{rf}$. By decreasing $I_{rf}$,
$\Delta Z_{real}$ first increases and then levels off at low
$I_{rf}$. Thus, $\Delta Z_{real}$ is "ohmic" at low $I_{rf}$,
consistent with the Fiory experiment \cite{Fiory} and theory
\cite{SH}.

Next we turn to the voltage broadening $\delta V$ of the rounded
jump. Figure 2 shows the rf current dependence of $\delta V$. We
determine $\delta V$ by taking the full width at half maximum of
the peak in the differential plot. It is clear that, with reducing
$I_{rf}$, $\delta V$ increases monotonically and seems to diverge
at $I_{rf}\cong$0. Since $\Delta Z_{real}$ is constant, this
indicates that the jump is smeared out toward $I_{rf}\rightarrow
0$.

Such divergence of $\delta V$ implies a pronounced spread in the
internal frequency, i.e., a wide distribution of local velocities
and/or lattice spacings. In fact, we find that no narrow band
noise at $f_{int}$ is detectable in the dc vortex flow. With
increasing $I_{rf}$, however, the spread in $f_{int}$ gradually
decreases. This can be explained by the frequency pulling effect
where the distribution of $f_{int}$ narrows and becomes locked
into $f$ \cite{NBN}. Thus, with increasing rf drive, the  ML
gradually enhances the order of the moving vortex lattice.

At high rf drives, the ML jumps in the $Z_{real}(V)$ curves
disappear. Therefore, we determined $\delta V$ from the measured
current steps in the dc $I-V$ curves. The open symbols in Fig.2
show that this $\delta V$ first decreases monotonically and then
levels off at large $I_{rf}$. This indicates that complete
(characterized by $\delta V$=0) ML does not occur, even at high rf
drives. Since $\delta V$ is more than one decade larger than what
is expected for purely elastic deformations \cite{Fiory,SH}, we
conclude that slip between moving rows and possibly remaining
plastic regions in the channel partially influence the ML
characteristics. This plasticity is caused by the disordered
vortex configurations in the channel edges.

\begin{figure}
\begin{center}
\epsfig{file=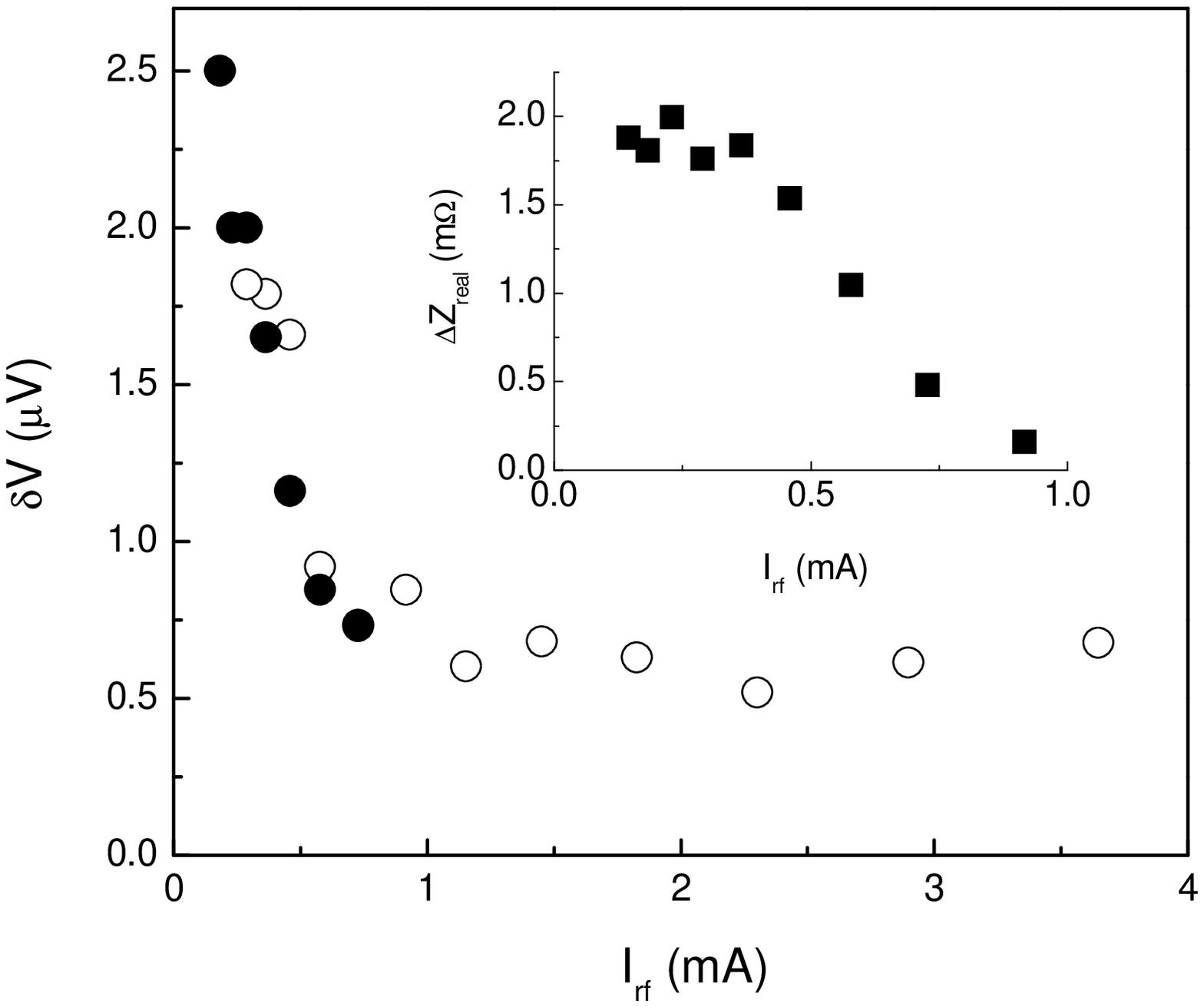, width=5.5cm} \vspace{0.1cm}
\label{Fig.2} \caption{Plot of voltage widths for fundamental jump
in real part of rf impedance (solid symbols) and fundamental dc
current step (open symbols) as a function of rf current. The
height of the impedance jump is plotted against rf current in the
inset.}
\end{center}
\end{figure}

This work was supported by the 'Stiching voor Fundamenteel
Onderzoek der Materie' (FOM).

\end{multicols}

%\clearpage
%\newpage
%\ FIGURE CAPTIONS
\end{document}